\documentclass[aps,prb,twocolumn,showpacs,amsmath,superscriptaddress]{revtex4}
\usepackage{graphicx}
\usepackage{bm}

\begin{document}

\title{Proximity Effects and Nonequilibrium Superconductivity in Transition-Edge Sensors}

\author{John E. Sadleir}
\affiliation{%
Department of Physics, University of Illinois, 1110 West Green Street, Urbana, IL 61801-3080}
\affiliation{%
   NASA Goddard Space Flight Center, 8800 Greenbelt Road, Greenbelt, MD 207701}
\email[]{john.e.sadleir@nasa.gov}

\author{Stephen J.\ Smith}
\affiliation{%
   NASA Goddard Space Flight Center, 8800 Greenbelt Road, Greenbelt, MD 207701}
\affiliation{%
CRESST and University of Maryland Baltimore County, MD 21250}

\author{Ian K.\ Robinson}
\affiliation{%
  London Centre for Nanotechnology, University College, London WC1E 6BT}
\affiliation{%
Department of Physics, University of Illinois, 1110 West Green Street, Urbana, IL 61801-3080}

\author{Fred M. Finkbeiner}
\affiliation{%
   NASA Goddard Space Flight Center, 8800 Greenbelt Road, Greenbelt, MD 207701}

\author{James A. Chervenak}
\affiliation{%
   NASA Goddard Space Flight Center, 8800 Greenbelt Road, Greenbelt, MD 207701}

\author{Simon R. Bandler}
\affiliation{%
   NASA Goddard Space Flight Center, 8800 Greenbelt Road, Greenbelt, MD 207701}
\affiliation{%
CRESST and University of Maryland College Park, MD 20742}

\author{Megan E. Eckart}
\affiliation{%
   NASA Goddard Space Flight Center, 8800 Greenbelt Road, Greenbelt, MD 207701}
\affiliation{%
CRESST and University of Maryland Baltimore County, MD 21250}

\author{Caroline A. Kilbourne}
\affiliation{%
   NASA Goddard Space Flight Center, 8800 Greenbelt Road, Greenbelt, MD 207701}

\date{submitted: May 11, 2011}

\begin{abstract} 

We have recently shown that normal-metal/superconductor (N/S) bilayer TESs (superconducting Transition-Edge Sensors) exhibit weak-link behavior.\cite{Sadleir_prl_10}  
Here we extend our understanding to include TESs with added noise-mitigating normal-metal structures (N structures).  We find TESs with added Au structures also exhibit weak-link behavior as evidenced by exponential temperature dependence of the critical current and Josephson-like oscillations of the critical current with applied magnetic field.  We explain our results in terms of an effect converse to the longitudinal proximity effect (LoPE)\cite{Sadleir_prl_10}, the lateral inverse proximity effect (LaiPE), for which the order parameter in the N/S bilayer is reduced due to the neighboring N structures.  Resistance and critical current measurements are presented as a function of temperature and magnetic field taken on square Mo/Au bilayer TESs with lengths ranging from 8 to 130 $\mu$m with and without added N structures.  We observe the inverse proximity effect on the bilayer over in-plane distances many tens of microns and find the transition shifts to lower temperatures scale approximately as the inverse square of the in-plane N-structure separation distance, without appreciable broadening of the transition width.  We also present evidence for nonequilbrium superconductivity and estimate a quasiparticle lifetime of $1.8\times10^{-10}$ s for the bilayer.  The LoPE model is also used to explain the increased conductivity at temperatures above the bilayer's steep resistive transition.

\end{abstract}

\maketitle

\section{\label{intro}Introduction}

\begin{figure}[ht]
\includegraphics[width=8.5cm]{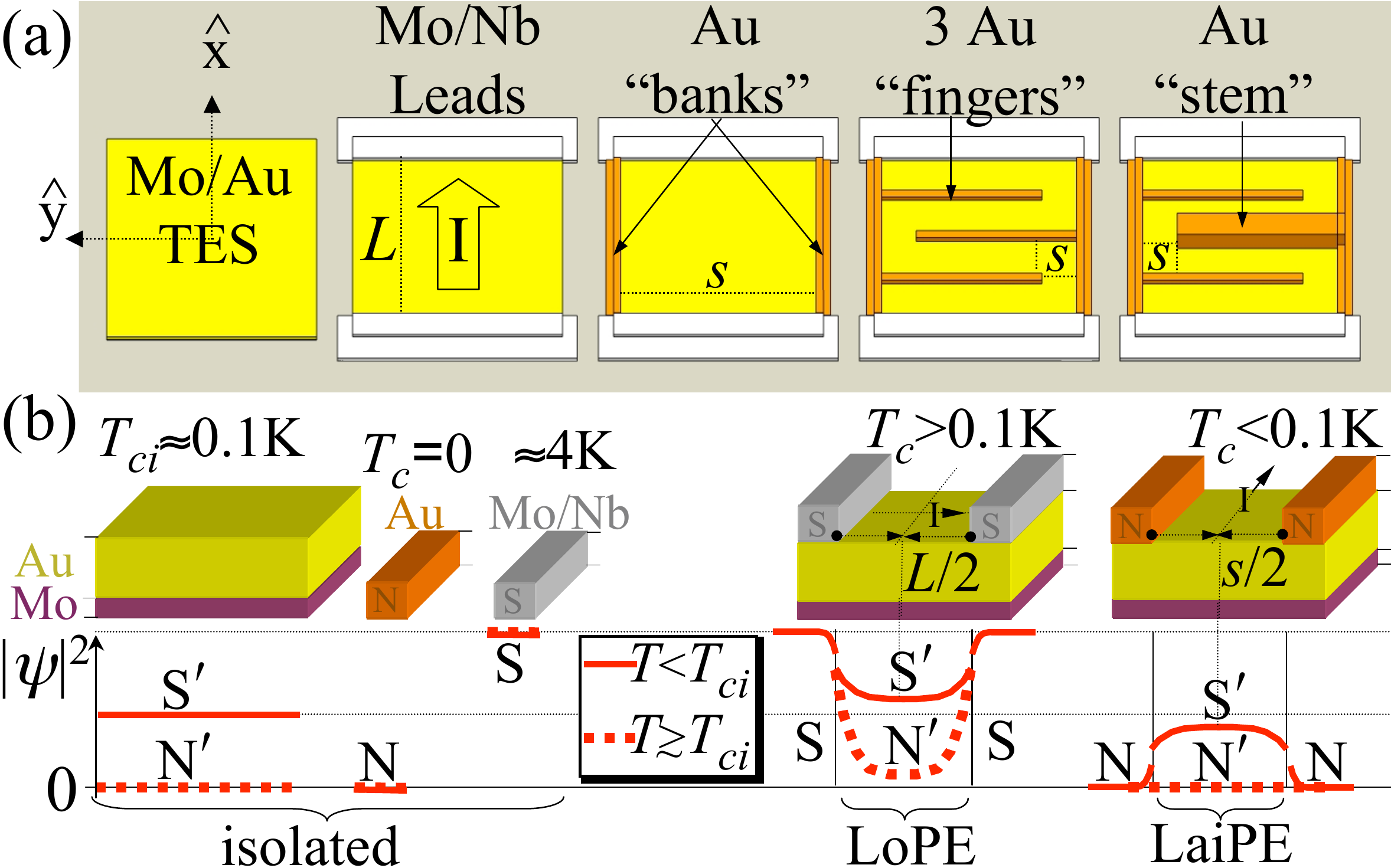}
\caption{%
(a) Schematic of TES sensors (color available online).  Square Mo/Au bilayer with attached Mo/Nb leads.  The current flows from lead to lead and the lead-separation distance is defined as $L$.  Au ``banks'' are added to prevent Mo shorts along the edge, Au ``fingers'' are added to reduce the unexplained electrical noise, and Au ``stems'' are added to provide attachment points for x-ray absorbers.  The minimum added Au structure separation distance is defined as $s$.  (b) In-plane variation of $|\psi|^2$ plotted for $T<T_{ci}$ (solid red curves) and for $T\gtrsim T_{ci}$ (dotted red curves) underneath the respective structures in isolation on the left (Mo/Au bilayer, Au, and Mo/Nb) and coupled heterostructures on the right (LoPE and LaiPE).  For a bilayer the average superconducting pair density $|\psi|^2$ and average $T_c$ is uniform across the wafer.  When higher $T_c$ Mo/Nb leads are attached the order parameter strength is increased above the average near the leads and decays with distance away from the leads to a minimum $L/2$ away (LoPE).  When Au structures are added, for $T<T_{ci}$ the order parameter strength is depressed near the structures and increases to a maximum $s/2$ away (LaiPE).
}
\label{schematic}
\end{figure}

	Superconducting Transition-Edge Sensor (TES) microcalorimeters\cite{IrwinHilton05} have been developed with measured energy resolutions in the x-ray and gamma-ray band of 
$\Delta E=1.8\pm$0.2 eV FWHM at 6 keV, \cite{Bandler08} 
and 
$\Delta E=22$ eV FWHM at 97 keV, \cite{Bacrania09}
respectively--- with the latter result at present the largest reported $E/\Delta E$ of any non-dispersive photon spectrometer.   In both examples the TESs are made of normal-metal/superconductor (N/S) proximity-coupled bilayers.  The TESs in both examples also have additional normal-metal interdigitated fingers (see Fig.\ \ref{schematic}a) which are found empirically to reduce unexplained noise\cite{Ullom04}.  
A complete theoretical understanding of the TES resistive transition including unexplained resolution-limiting noise sources and how the added N structures change the TES is desired to help guide this exciting technology to its full potential.  

There is also a renewed interest in understanding S-N heterostructures more generally.\cite{Cuevas07,Heersche07,leSueur08}  
Driven in part by advances in fabrication capabilities, improved understanding of S-N interactions is motivating new superconducting device concepts.  The richness of physics arising from S-N heterostructures is considerable and with potential applications including improved magnetic sensing, nanocoolers, particle detection, THz electronics, and superconducing qubits. \cite{Saira07,Giazotto10,Jarillo-Herrero06}

Previous attempts to model the TES resistive transition include using Kosterlitz-Thouless-Berezinski (KTB) theory,\cite{Fraser04} fluctuation superconductivity,\cite{Seidel04} percolation theory for a random superconducting resistor network,\cite{LindemanPerc06,Brandt09}  and thermal fluctuation models.\cite{Luukanen03}  We have recently shown both experimentally and theoretically that TESs exhibit weak-link behavior, where, unlike previous models, the average order parameter varies over the TES.\cite{Sadleir_prl_10}  Here, we present further evidence for a spatially varying order parameter over the TES, and extend our understanding to TES devices with added N structures.  We now show that our measurements of the transition have a natural explanation in terms of a spatially varying order parameter that is enhanced in proximity to the higher $T_c$ superconducting leads (the longitudinal proximity effect or LoPE)\cite{Sadleir_prl_10} and suppressed in proximity to the added N structures (the lateral \emph{inverse}\cite{Buzdin05} proximity effect or LaiPE) as depicted in Fig.\ \ref{schematic}b.

In Ref. [\onlinecite{Sadleir_prl_10}] we showed that the higher $T_c$ superconducting leads enhance superconducting order longitudinally into the N/S bilayer over remarkably long lengths in excess of 100 $\mu$m, over 1000 times the mean free path.  Our theoretical model agreed with the critical current measured over 7 orders of magnitude versus temperature for square TESs ranging in size from 8 to 290 $\mu$m and over a factor of 3 change in the effective transition temperature.  We also showed that the temperature dependence of the critical current explains the measured resistive transition widths.  The transition temperature of the TES was found to scale linearly with the transition width and both scale approximately as $1/L^2$, where $L$ is the lead separation.  
In this GL model the longitudinally lead/bilayer/lead TES is treated as an S/N$^\prime$/S or S/S$^\prime$/S structure for temperatures $T$ above or below 
respectively the intrinsic transition temperature of the bilayer in isolation $T_{ci}$, with $T_{cL}$ is the transition temperature of the leads satisfying $T_{cL}>T_{ci}$.  The model shows even for $L \gg \ell_{mfp}$ and $L \gg \xi_N$ and $T>T_{ci}$ the superconducting order parameter remains finite in the center of the TES (at $x=0$) for all $T<T_{cL}$; where $ \ell_{mfp}$ and $\xi_N$ are the mean-free path and normal metal coherence length of the TES.

In Sec. \ref{paraconductivity_section} we show that the longitudinal proximity effect also explains the significant change in resistance at temperatures above the abrupt resistance change (\emph{i.e.}, at temperatures above the effective transition temperature $T_c$).  We also present measurements in Sec. \ref{RT_shifts_and_weak_link_behavior_section} and derive in Sec. \ref{laipe_scaling_section} that the effect of adding additional normal-metal structures shifts $T_c$ to lower temperature by an amount that scales approximately as $1/s^2$, where $s$ is the normal-metal structure separation distance.

The regions with added N structures (N/N/S layer regions)  suppress superconducting order laterally into the N/S bilayer, Fig.\ \ref{schematic} (b).  The TES designs giving the best energy resolution,\cite{Bandler08,Bacrania09} having both leads and added N structures, have a spatially varying order parameter due to these two competing effects--- superconducting enhancement in proximity to the leads and suppression in proximity to the added N structures.  In both cases the spatially varying order parameter means that the transition temperature for the TES is an effective transition temperature because it is highly current dependent.  The critical current $I_c$ (the current at which superconductivity first breaks down) depends exponentially upon the square root of the temperature $T$ and the lengths $L$ and $s$ of the weak-link TES.\cite{Sadleir_prl_10}  In this theoretical framework the first onset of resistance occurs when the TES current reaches a local critical current density $j_c$ for the minimum (maximum) order parameter along series (parallel) connected regions.  
In Sec. \ref{IcB_Kxy_section} a new variant of Josephson interferometry is used showing the spatial variation of the lateral inverse proximity effect with distance from added Au structures.  In addition to the lateral inverse proximity effect we show in Sec. \ref{nonequlibrium_sc_section} that the added Au structures also introduce charge imbalance or nonequilibrium superconductivity.

\section{Samples and Measurements}
\label{samples_section}
Our TESs consist of a 45 to 55 nm thick Mo layer ($T_c \approx$ 0.9 K) to which 190 to 210 nm of Au is added giving a combined intrinsic bilayer transition temperature $T_{ci} \approx$ 100 to 170 mK.  The added Au structures in the form of fingers and/or banks are electron-beam evaporated with a 350 nm thickness and the Au stems are electrodeposited to a thickness of 1 to 4 $\mu$m.  We estimate the intrinsic transition temperature for a trilayer of Mo/Au/Au of thickness 50/200/350 nm to be about 5 mK, and lower for the thicker stem trilayer.\cite{Martinis00}  This means that fingers, banks, and stem Mo/Au/Au trilayer structures in isolation would be normal metal at all temperatures above 5 mK. The TES is connected to Mo/Nb  leads with intrinsic values of $T_c$ $\approx$ 3 to 8 K.  We find that temperatures much larger than used in device fabrication are needed to cause measurable (by x-ray diffraction or energy dispersive spectroscopy) interdiffusion between the Nb, Mo, Au systems, ruling out interdiffusion at interfaces as an explanation of the results.\cite{Sadleir_phd_10,stress_comment}
Further details on the device fabrication process, device electronics, and measurement techniques used can be found in Refs.\ \onlinecite{Sadleir_prl_10, Chervenak04}.  

\section{Transition Shifts from S and N Structures \& Weak-Link Behavior}
\label{RT_shifts_and_weak_link_behavior_section}
Measurements of the TES resistance $R$ are made by applying a sinusoidal current of frequency 5-10 Hz and
amplitude $I_{bias} \sim$ 50-250 nA, with zero dc component, to the TES in parallel with a 0.2 m$\Omega$ shunt resistor ($R_{sh}$). The time-dependent TES current is measured with a SQUID feedback circuit with input coil in series with the TES.  

\begin{figure}[h]
\includegraphics[width=8.5cm]{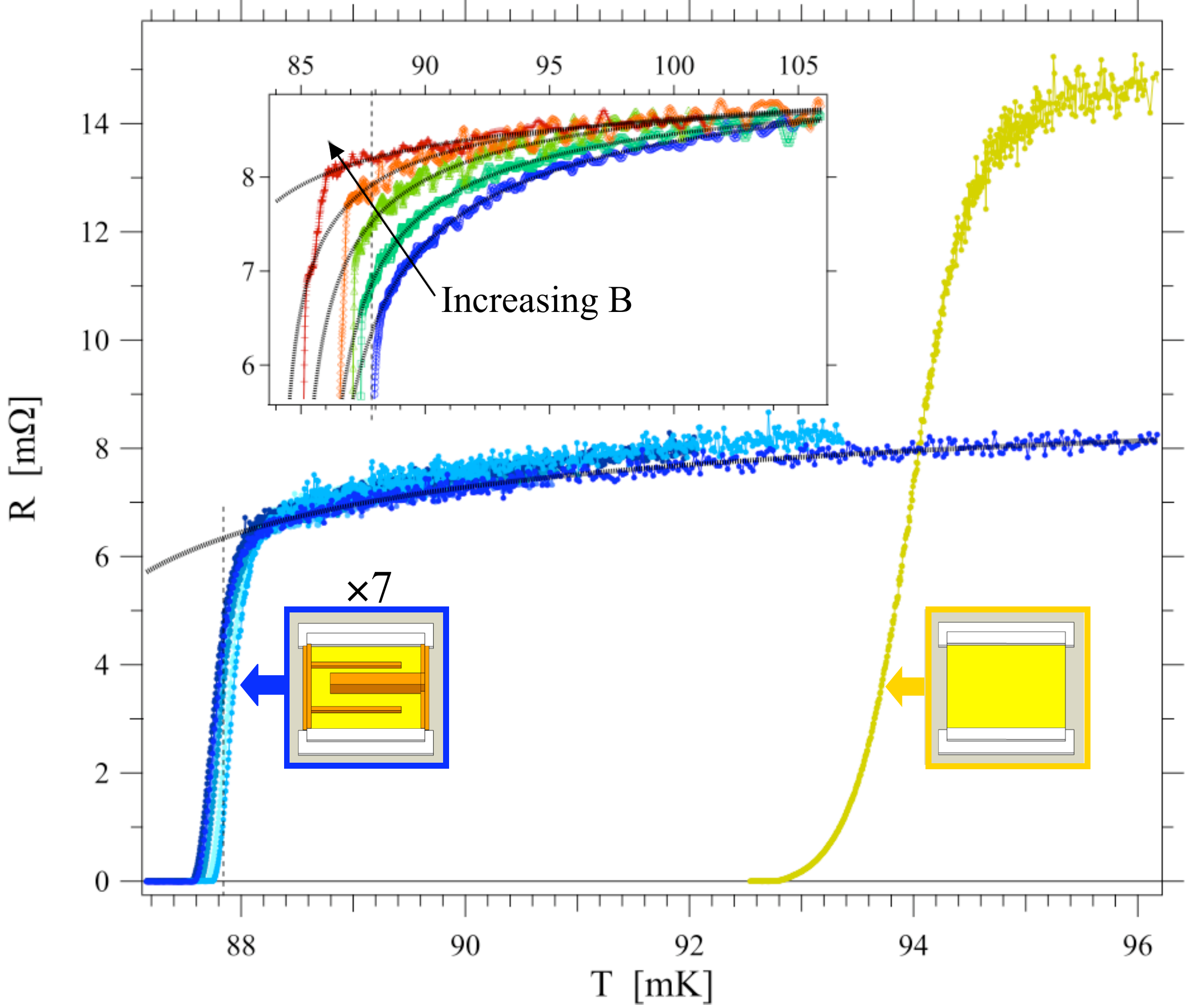}
\caption{%
RT measurements of 7 devices ($L = $ 110 $\mu$m) of the same design from the same Mo/Au bilayer deposition ($T_{ci} \approx $ 93 mK)  with Au banks, fingers, and absorber stem showing good uniformity and a $T_c$ reduction of $\approx$ 6 mK compared to a neighboring device with no added Au structures; clear evidence of the lateral inverse proximity effect of the added Au structures reducing $T_c$ in the Mo/Au bilayer. The inset shows R(T) data in an applied magnetic field (0, 27, 41, 57, 80  mG). 
The dotted curves are fits to the higher resistance Ònormal-stateÓ RT region using the longitudinal proximity effect model described in the text (Eq. \ref{N_slope}).  The vertical dashed line is the $T_c$ for $B=0$ defined at $R = R_N/2$.
}
\label{rt_uniformity}
\end{figure}

RT (resistance versus temperature) measurements from seven pixels of identical design were performed using an array showing good uniformity,\cite{Smith08} average $T_c$ defined at $R = 0.5 R_N$ of 87.84 mK $\pm$ 0.06 mK, and a full range spanning 0.16 mK as shown in Fig.\ \ref{rt_uniformity}.  The average transition width between $R$=1.0 m$\Omega$  and $R$ = 5.0 m$\Omega$ for the seven pixels is 0.194 $\pm$ 0.011 mK. The reduction in $T_c$ due to the added Au structures is explained in Sec. \ref{laipe_scaling_section} in terms of lowering the order parameter in Mo/Au laterally a distance $s/2$ away.  Note the effect of the added Au structures shifts the entire transition to lower temperatures and the size of the shift is much larger than the spread in the measured $T_c$ from pixels of the same design. 

\begin{figure}[h]
\includegraphics[width=8cm]{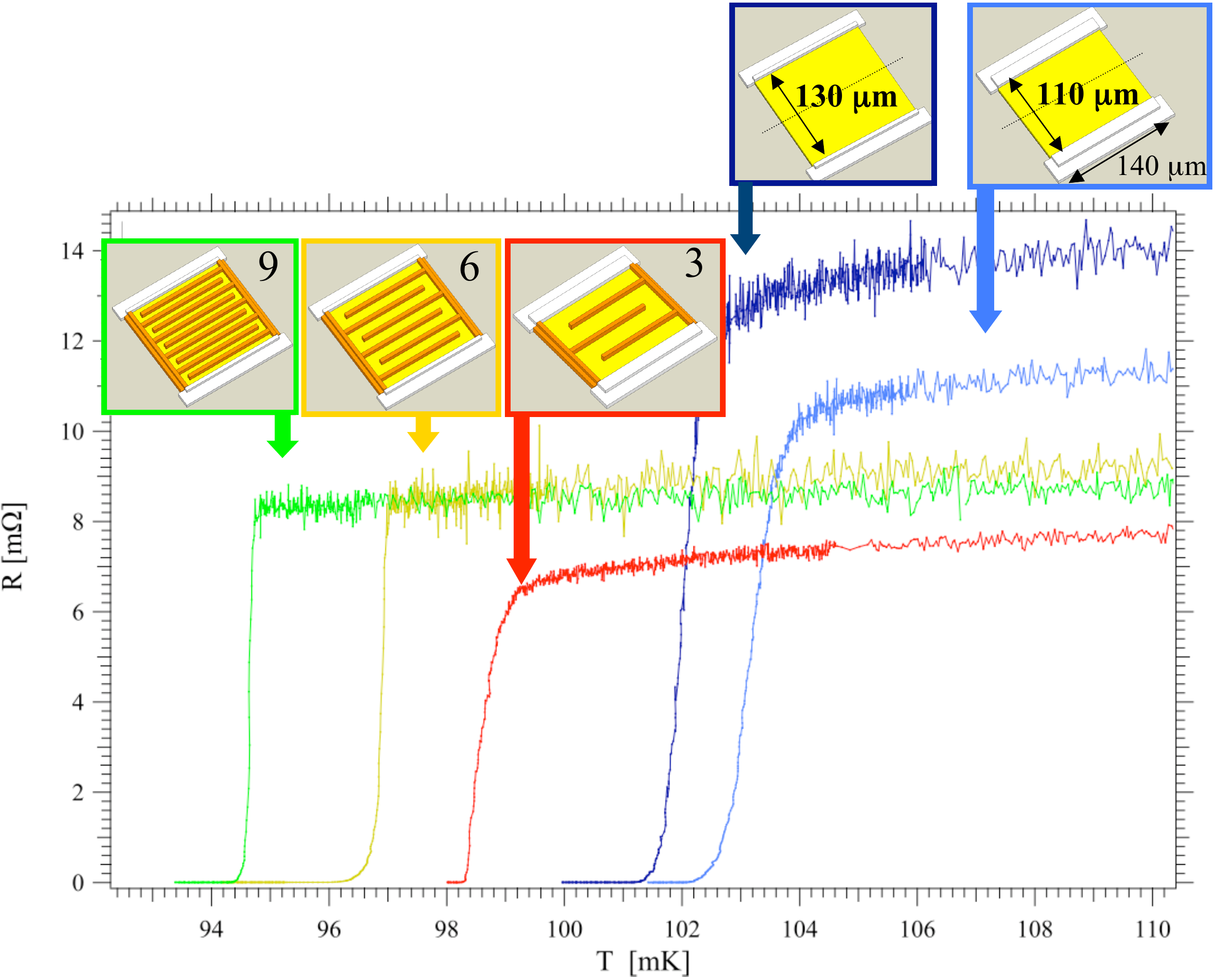}
\caption{%
RT measurements of 5 different devices from the same Mo/Au bilayer deposition ($T_{ci} \approx $ 101 mK).  Mo/Au bilayers with $L$ = 110 and 130 $\mu$m show a small shift in the effective transition temperature consistent with the longitudinal proximity effect model.  The devices without added Au structures are compared to devices with $L$ = 110 and 130 $\mu$m and 3, 6, and 9 interdigitated fingers of additional Au of 350 nm thickness and 5 $\mu$m width showing their effect is to shift the entire resistive transition to lower temperatures.
}
\label{RT_fingers}
\end{figure}

In Fig. \ref{RT_fingers} \; RT measurements of 5 different device designs with the same bilayer composition are shown.  The two TESs with no added Au structures have a small difference in $T_c$ consistent with the different $L$ values through the longitudinal proximity effect.  The added Au structures shift the entire resistive transition to lower temperatures, with the size of the shift increasing with number of fingers.  The additional normal-metal structures produce parallel paths for current to flow and lower the normal-state resistance (Figs. \ref{rt_uniformity} and \ref{RT_fingers}) consistent with a resistor network model including the geometry and measured resistivities of each layer in isolation.\cite{Rn3finger}

\begin{figure}[h]
\includegraphics[width=8cm]{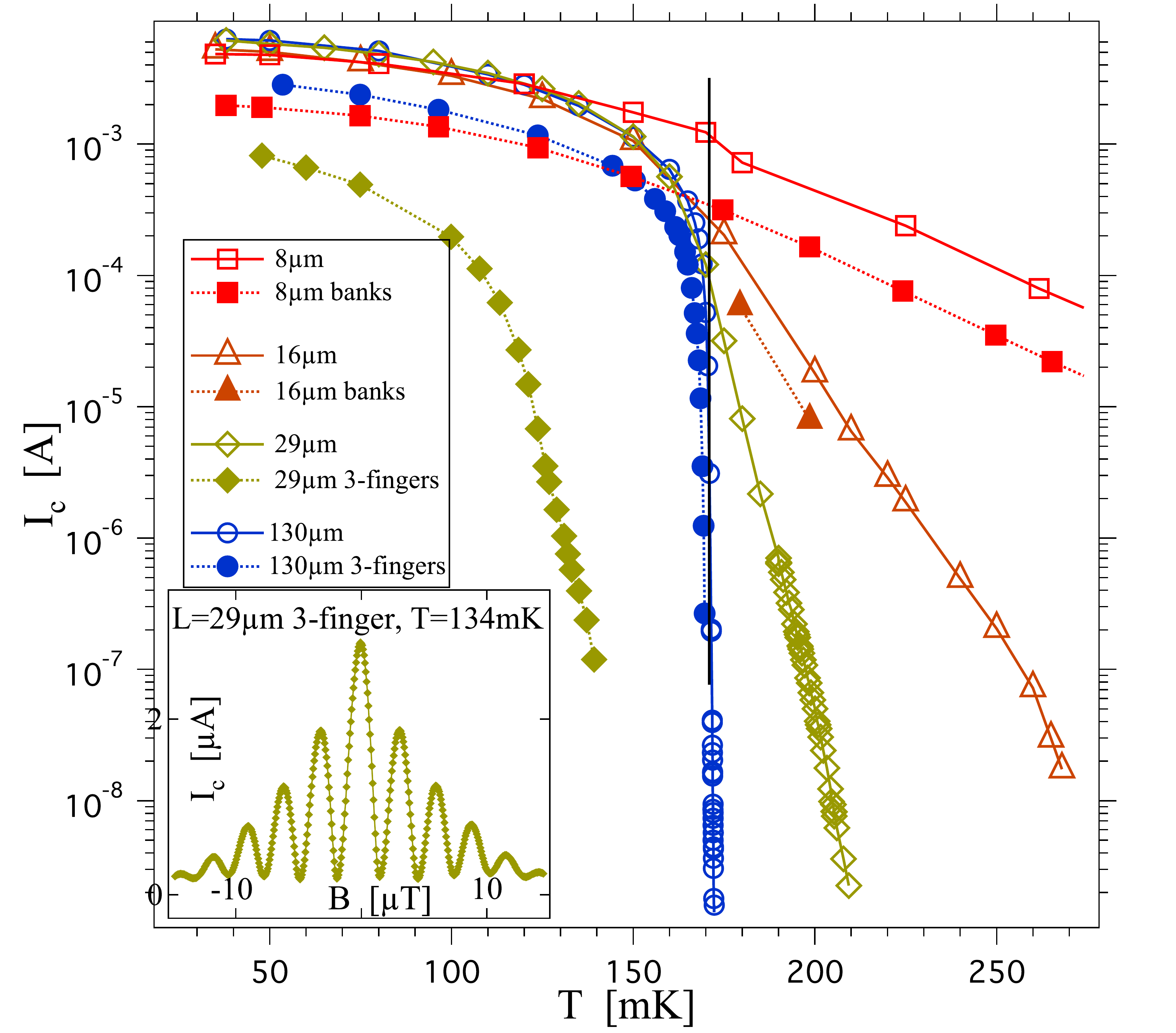}
\caption{%
Comparison of measured critical current versus temperature for devices with and without added Au structures for L ranging from 8 to 130 $\mu$m.  Vertical line is the intrinsic transition temperature of the Mo/Au bilayer $T_{ci}\approx 171$ mK (no LoPE or LaiPE).  Open markers 
are for square devices with no added Au structures (LoPE only).  Solid markers 
are for square devices with three Au fingers and/or Au banks along the edges (LoPE and LaiPE).  
Inset shows clear Josephson-like oscillations of the critical current with applied magnetic field.  
}
\label{Ic_T}
\end{figure} 
This temperature shift of the transition from adding Au structures is also seen in measurements of the critical current $I_c(T)$ as a function of temperature in Fig. \ref{Ic_T}.  Notice the impact of adding Au structures for a device of the same $L$ is that the $I_c(T)$ transition is shifted to lower temperatures with approximately the same exponential decay constant with temperature, and therefore approximately the same transition width.   For the $L$ = 29 $\mu$m device adding Au banks and fingers causes a transition shift of nearly 60 mK with the same exponential temperature decay as the $L$ = 29 $\mu$m device with no added Au structures. This characteristic is seen in devices ranging from $L=8$ to 130 $\mu$m.
 
In addition to the exponential temperature scaling of the critical current, further evidence for weak-link behavior is shown in the inset of Fig. \ref{Ic_T}.  The critical current versus applied magnetic field data collected for the $L$ = 29 $\mu$m device with three Au fingers exhibits Josephson-like oscillations of the critical current (inset of Fig. \ref{Ic_T}).  The oscillation period $\delta B$ implies an effective area $\Phi_0/\delta B\approx$ 795 $\mu$m$^2$, where $\Phi_0$ is the magnetic flux quantum.  The $I_c(B)$ pattern indicates that despite a large fraction of the TES bilayer covered with the Au interdigitated fingers the entire TES is acting as one coherent weak-link.

\section{$T_c$ and $\Delta T_c$ Change with Au Fingers Inconsistent with Percolation Model}
\label{percolation_section}

Percolation models have garnered attention in recent years in efforts to explain the resistive transition in TESs.\cite{LindemanPerc06,Brandt09}  It was hypothesized that added N fingers changed the TES transition behavior by altering the geometry of the percolating paths between electrodes. In this section we apply a percolation analysis to our TES devices and compare with measurements of the resistive transition.  We present a local geometric argument for why the $T_c$ could decrease with increasing number of fingers, but find that other characteristics are inconsistent with our measurements.

Suppose  our square Mo/Au bilayer is made up of a rectangular grid with bin dimensions $\delta x$ and $\delta y$ both small relative to the length $L$ and width $W$.  Suppose each bin has a local characteristic $T_c$ and the distribution of $T_c$s for all bins is approximately Gaussian distributed.  We could imagine this random local $T_c$ distribution arises from variations in the local Mo or Au thickness, Mo/Au interface transmissivity, film stress, impurities, defects, etc.  We represent the TES as a two-dimensional random superconducting resistor network such that at any temperature $T$ there will be a fraction of domains $p$ that are superconducting with zero resistance, and a fraction of domains that are normal $(1-p)$ with finite resistance.  As the temperature is lowered the concentration of superconducting sites or bonds increases and the measured resistance of the network decreases.  The resistance of the network is zero when a continuous percolating path of superconducting domains spans $x=-L/2$ to $x=L/2$.  

We can now see how with the same distribution of local $T_c$s  the measured $T_c$ of the network can shift with changes in aspect ratio.  A Mo/Au bilayer geometry longer in length along $x$ (width along $y$) will on average require a higher (lower) concentration of superconducting domains for the same network resistance fraction and the network $T_c$ will be shifted to lower (higher) temperatures.  If adding the Au fingers is thought of as effectively increasing the length to width ratio for a meandering path around the fingers then the network $T_c$ will be shifted to lower temperatures.  But in this model the $T_c$ shift to lower temperatures would also be accompanied by an increase in transition width.  Figures \ref{rt_uniformity} and \ref{RT_fingers} show that adding Au fingers and/or stems shifts the transition to lower temperatures by amounts many times the transition width $\Delta T_c$ and without increasing $\Delta T_c$.  

We conclude that our measurements are inconsistent with a local model causing the shifts in transition temperature.  Our measured $I_c(T), I_c(B)$, and resistive transition is explained in terms of nonlocal coherence effects, whereby superconducting correlations in Mo/Au bilayer are altered by Mo/Nb and added Au structures over lengths many times the electron mean free path.  

\section{Longitudinal Proximity Effect Conductivity Enhancement for $T > T_c$}
\label{paraconductivity_section}

At temperatures well above a uniform superconductor's $T_c$ the sample is in the normal state and has an associated normal-state resistance $R_N$ that has a weak temperature dependence associated with the normal metal at low temperatures.  

The RT measurements in Fig. \ref{rt_uniformity} show that at temperatures above the abrupt resistive transition the resistance is not constant and has nonzero slope.  The inset of Fig. \ref{rt_uniformity} shows that when a uniform applied magnetic field along the film thickness direction increases (0, 27, 41, 57, 80 mG) the abrupt drop in resistance is shifted to lower temperatures.  In addition, as the magnetic field is increased the size of the enhanced conductivity at temperatures above the abrupt change in resistance decreases.  For $T>T_c$ a 27\% reduction in resistance (blue curve) for $B=0$ is reduced to a 5\% resistance change for $B$ = 80 mG (red curve).  Similar enhancements in conductivity for $T>T_c$ are seen in other samples. 

We first consider superconducting fluctuations as a possible explanation.   Excess conductivity mechanisms (also called paraconductivity) in a superconductor near the transition has been studied and experimentally confirmed for some time.\cite{Glover67}  Originally Ginzburg demonstrated that in clean bulk superconductors fluctuation phenomena only becomes important in the very narrow temperature region ($\sim10^{-12}$ K) about $T_c$.\cite{Glover67}  It was later demonstrated by Aslamazov and Larkin for dirty superconducting films that fluctuations are determined by the conductance per square and could be important over much wider temperature ranges than bulk samples.\cite{Aslamazov68}  For a uniform superconductor at temperatures above the superconducting phase transition superconducting pair fluctuations lowers the resistivity below its normal-state value.  We find that Aslamazov-Larkin (AL) fluctuations \cite{Tinkham06}  predict a transition drop to $R=0.90\,R_N$ that is $\sim$$\mu$K above $T_c$, whereas the normal-state slope in Fig. \ref{rt_uniformity} is orders of magnitude larger $\sim$ mK.  In a clean superconductor the Maki-Thompson (MT)\cite{Maki68} term can be as much as an order of magnitude larger than the AL contribution but still another physical explanation is needed to explain our measurements. \cite{Skopol75}  Seidel and Beleborodov \cite{Seidel04} calculated the fluctuation superconductivity resistive transition width in TES sensors and also found the calculated widths to be orders of magnitude smaller than the measured values.

We next show that the enhanced conductivity above the abrupt phase transition has a natural explanation in terms of the longitudinal proximity effect. 
The characteristic length over which superconducting order will penetrate into a metal is given by the normal-metal coherence length.\cite{Tinkham06,DeGennes99}  The sloped normal-state region is fit to a $1/\sqrt{T-T_{c\,N'}}$ temperature scaling assuming the zero-resistance region penetrates longitudinally a distance of twice the normal-metal coherence length into the TES from each lead and is normal beyond.  We may then express the temperature dependence of the resistance above the abrupt transition as 
\begin{equation}
R(T)=\frac{dR}{dx} \left[ L-4\;\sqrt{\frac{\hbar v_F \ell_{mfp}}{6 \pi k_B (T-T_{c\, N'})}} \;\right]
\label{N_slope}
\end{equation}
where $T_{c\, N'}$ is a fit parameter corresponding to the effective transition temperature.   Including the reduction in resistance from the Au banks and width of the device in Fig.\ \ref{rt_uniformity} the resistance per length $\frac{dR}{dx}\approx$ 92 $\Omega$/m.  The Mo/Au bilayer has a measured normal-state resistance $\approx$ 20 m$\Omega$/sq.  Including the carrier density ($5.9\times10^{28}$ 1/m$^3$) and Fermi velocity ($v_F=1.39\times10^6$ m/s) for Au we find the mean free path is thickness limited, $\ell_{mfp}$ = 210 nm and the electronic diffusivity $D$ = 0.0968 m$^2/$s.  The black curve in Fig. \ref{rt_uniformity} is Eq. \ref{N_slope} with these parameter values showing agreement with the data for $T>T_c$.

The data in the inset of of Fig. \ref{rt_uniformity} is also fit using equation Eq. \ref{rt_uniformity} with the same set of parameter values above held constant while letting $T_{cN'}$ and $\ell_{mfp}$ decrease monotonically with increasing $B$, reaching $T_{cN'}=78.5$ mK and $\ell_{mfp}=90$ nm at the largest $B$.  This may be qualitatively understood in terms of a modest magnetic field increase causing an increase in spin-flip scattering and a reduction in the depth of the lead induced minigap's penetration into the bilayer.

The RT curves of Fig. \ref{RT_fingers} with 6 and 9 Au fingers added show much less enhanced conductivity for $T>T_c$.  This is also consistent with the longitudinal proximity effect interpretation because in this case the propagation of the diffusing superconducting order from the leads (LoPE) is opposed by the converse effect from the Au fingers (LaiPE) and as a result there is less resistance change for temperatures above $T_c$.

\section{Lateral Inverse Proximity Effect Scaling}
\label{laipe_scaling_section}

We have shown how the measured effective $T_c$ of a TES is a function of the lead separation L.\cite{Sadleir_prl_10}  We now show the effective $T_c$ of the Mo/Au bilayer is lowered by the addition of extra Au layer structures laterally many tens of microns away and how this change in $T_c$ scales with the added Au structures separation.  

We model our system as N/S/N structures corresponding to N regions of Mo/Au/Au and S regions of Mo/Au.  We follow the theoretical approach used by Liniger\cite{Liniger93} on an N/S/N sandwich using a one-dimensional nonlinear Ginzburg-Landau (GL) model where the length of the S layer is a variable.  Liniger showed that the GL order parameter vanishes if the length of the S layer, $s$, is less than a critical length $s_c$, where
\begin{equation}
s_c = 2\;\xi_{GL} (T)\; \arctan \left\{ \frac{\xi_{GL} (T)}{b}\right\}
\label{scGL}
\end{equation}
with the coherence length in the S layer $\xi_{GL}(T) = \xi_S(T)=\xi_S(0)/\sqrt{(1-T/T_c)}$ and $b$ the extrapolation length of the superconducting order parameter into the normal metal.  For an insulating interface, $b$ is infinite and the critical length vanishes.  For clean interfaces the electron transmission coefficient is unity, characterized by no scattering centers at the interface and no Fermi-velocity mismatch between N and S.  The clean N/S interface condition is well met for our geometries allowing us to set the extrapolation length equal to the normal-metal coherence of the N region.  Using coherence length expressions for N and S in the dirty limit,
\begin{equation}
\xi_N= \sqrt{\frac{\hbar D_N}{2\pi k_B T}}
\label{coherenceN}
\end{equation}
 and 
\begin{equation}
\xi_S(0)= \sqrt{\frac{\pi \hbar D_S}{8 k_B T_c}}
\label{coherenceN}
\end{equation} 
we then have
\begin{equation}
s_c = \sqrt{
\frac{\pi \hbar D_S}{2 k_B \left(T_c-T\right)}     }\;\;\;  \arctan \left\{ \frac{\pi}{2}\; \sqrt{\frac{D_ST}{D_N(T_c-T)}}\right\}
\label{scfull}
\end{equation}
where $D_S$ and $D_N$ are the electronic diffusivities in the S and N layer respectively.
In the limit of $T$ near $T_c$, $\xi_S(T)$ diverges, and $s_c \approx \pi \,  \xi_S(T)$.  Applied to our structures near $T_c$ this means the change in transition temperature due to the additional Au structures scales like the separation of Au structures to the negative 2 power ($\sim1/s^2$). 

\begin{figure}[h]
\includegraphics[width=8cm]{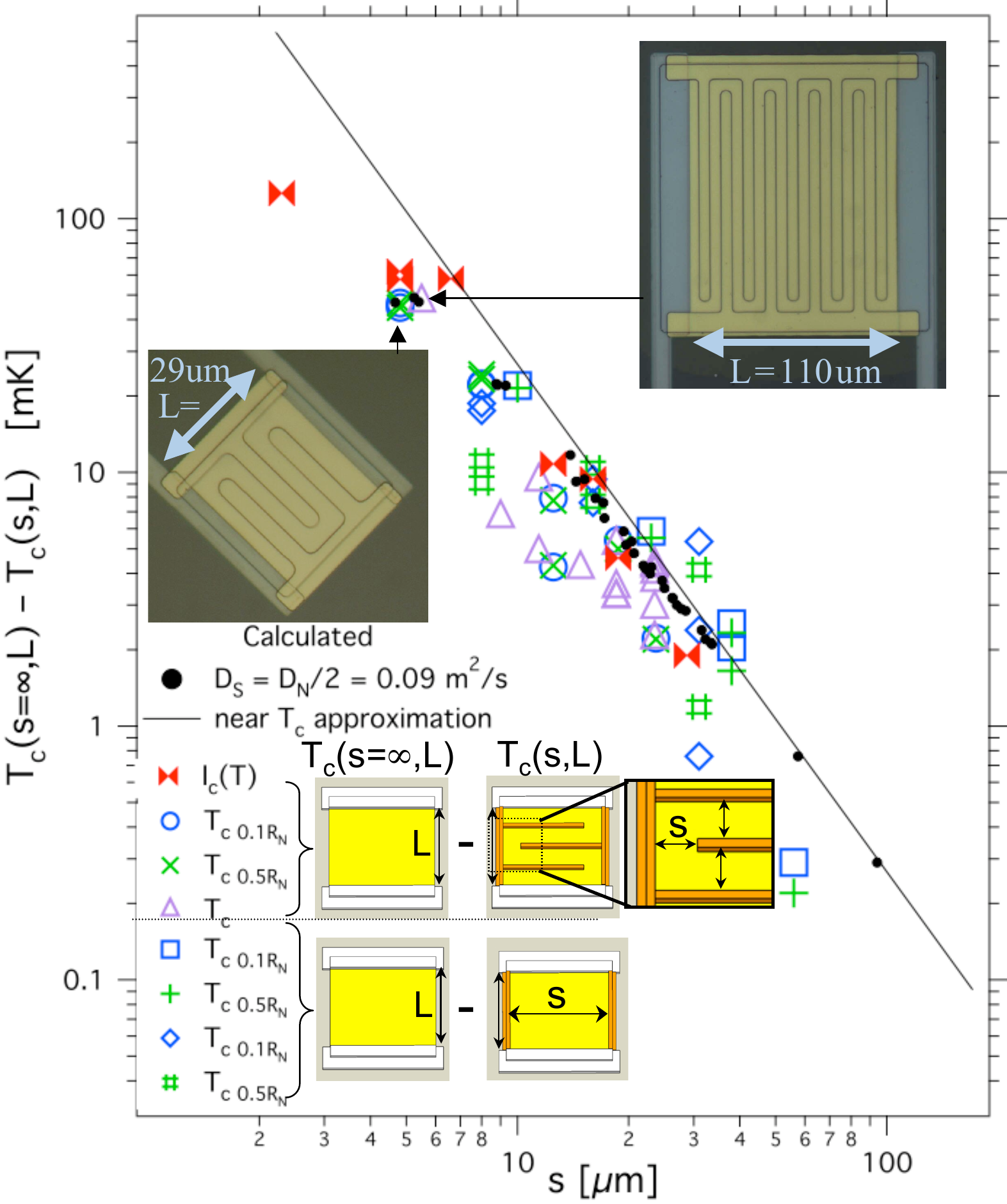}
\caption{%
Size of the shift in the effective $T_c$ for devices with added Au structures versus the separation distance $s$, from $I_c(T)$ measurements (red solid bow-tie markers) and all other markers from RT measurements.  The 40 samples cover many different bilayers and different pixel designs with different sizes $L$, separation $s$, number of Au fingers, and Au absorber stem structures.  
The black solid circles are calculated using Eq. (\ref{scfull}) with the superconductor and normal-metal diffusivities $D_S=0.09$ m$^2$/s and $D_N=2D_S$, consistent with the values determined in the text.  The solid line shows the dependence using the near $T_c$ approximation which departs from the solid circles at the smaller $s$ values.  Microscope pictures with arrows pointing to data points are shown for devices with very different $L$ values but similar $s$ values, both of which are consistent with Eq. (\ref{scfull}).  The arrow for the $L$ = 29 $\mu$m is pointing to the $\times$ and $\bigcirc$ markers, and the $L$ = 110 $\mu$m is pointing to the $\triangle$ marker.
}
\label{T_c_s}
\end{figure} 

Devices tested with added Au structures also have higher $T_c$ leads meaning there is a measured increase in $T_c$ from the leads and decrease in $T_c$ from the added Au structures.  Taking into account both the proximity effect of the leads and the inverse proximity effect of the added Au structures we plot the combined effect of devices tested over several years having many different added Au pattern structures, different TES sizes, over many different fabrication runs whenever $T_c$ for a device without added Au structures ($T_c(s=\infty,L)$) was measured and $T_c$ for a device from the same bilayer and same $L$ with Au structures added ($T_c(s,L)$).   We then plot $T_c(s=\infty,L)-T_c(s,L)$ in Fig.\ \ref{T_c_s}.  Both the $R=0.1R_N$ and $0.5R_N$ $T_c$ definitions exhibit similar scaling with $s$.  By comparing both $T_c$ definitions we also see how the added Au structures change the low current resistance measurements transition width.

Most of the pairs of points show that the $T_c$ shift for the $R=0.1R_N$ and $0.5R_N$ definitions are approximately equal which means the Au structure shifted the transition but did not change the transition width (i.e. the low current resistive transition width $\Delta T_c$ is mostly dependent upon $L$ as was found in square TESs with no added Au structures\cite{Sadleir_prl_10}).  There exist pairs of points that show a slight increase in $\Delta T_c$ and even a few with a slight decrease in $\Delta T_c$ upon adding the Au structures.   In Fig.\ \ref{T_c_s} we also plot the size of the temperature shift of the higher temperature $I_c(T)$ curve (red bow-tie markers) upon adding Au structures with separation $s$, exhibiting the same scaling of the $T_c$ shift  with $s$ as found by the RT measurements.  Similar consistency between the shift in $T_c$ from $I_c(T)$ and RT measurements was found for the longitudinal proximity effect  in  Ref. \onlinecite{Sadleir_prl_10} studying square devices with no added Au structures.

We find surprising agreement over the large diverse sample set using Eq. \ref{scfull} with $D_S$=$D_N/2$=0.09 m$^2$/s, consistent with the typical value of our electronic diffusivity for the Mo/Au determined from resistance measurements.  The scaling is observed over an $s$ range of 2.3 to 38 $\mu$m, a $T_c$ shift of over two decades, with the largest $T_c$ change from adding Au structures being 75\%, 23\%, and 37\% for multiple samples with $s$ of 2.3, 4.8, and 5 $\mu$m respectively.

We have found only one other report of an N/S/N system's $T_c$ scaling with size.  In Boogard {\it et al},\cite{Boogaard04} RT curves are taken for one-dimensional Al wires connected to Al/Cu normal-metal reservoirs.  Four of the five data points used for wire lengths of 2, 2.5, 3, and 3.5 $\mu$m are fit to the inverse square of the wire length with the largest change of $T_c$ being 8\%.

\section{LaiPE Spatial Variation of Critical Current Density from $I_c(B)$ Measurements}
\label{IcB_Kxy_section}

The resistance and critical current measurements are transport measurements that measure integrated properties over the samples' dimensions.  As a result, the critical current measurements probe the order parameter at a local minimum region in the sample.  We have shown in the previous sections how the strength of the order parameter for this region changes with current, temperature, distance from the S leads, and distance from additional N structures in a manner consistent with theory.

We are therefore still interested in a way to extend our LaiPE investigation to a measurement that is sensitive to and actually samples the spatial variation of the strength of the superconductivity.  In other words, we wish to directly measure the spatial variation of the superconducting order, in a single device, at a single temperature, showing superconductivity becoming suppressed as the added N structures are approached and increasing to a maximum value halfway between the added N structures.   Low temperature scanning tunneling microscopy (STM) studies have been carried out studying the gap in the density of states for S/N bilayers of variable thickness.\cite{Cretinon04}  They have also been used to study S islands coated with an N layer\cite{Moussy01} or N islands deposited on an S substrate.\cite{Tessmer96}  In both cases the STM measurements were made over lateral or longitudinal in-plane distances less than 20 nm away from the foreign body, much smaller than the relevant in-plane lengths in our system.   

We present a different technique to probe the spatial variation of the superconducting order in TESs in this section.  The critical current $I_c$ versus applied magnetic field $B$ is measured for devices with and without Au banks along the edges.  The analysis of the $I_c(B)$ extracts a spatially varying sheet current density for both samples showing how the addition of the Au banks changes the calculated critical current distribution in the TES.

\begin{figure}[h]
\includegraphics[width=8cm]{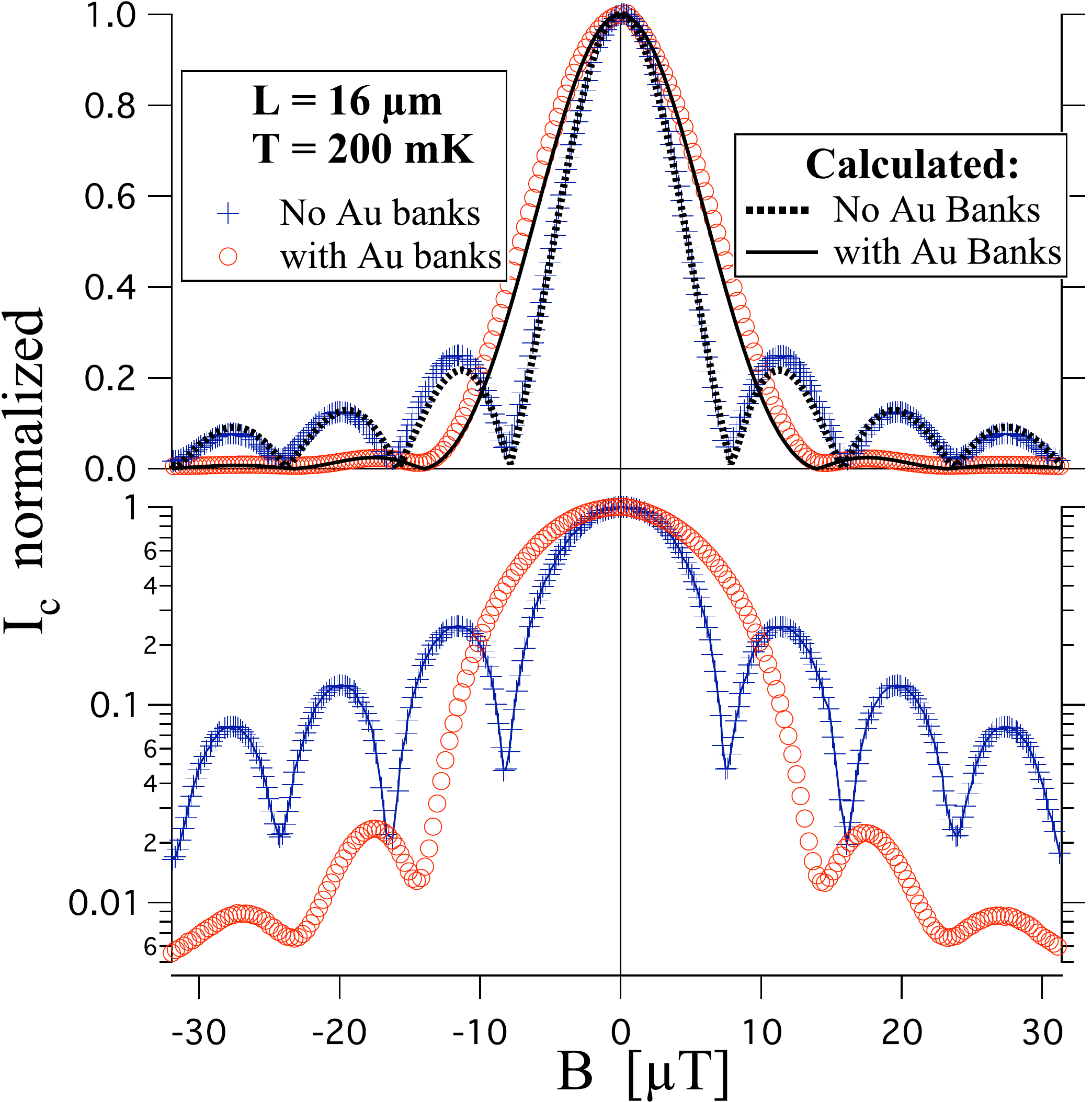}
\caption{%
Normalized critical current versus applied magnetic field measured for square $L=16\,\mu$m devices with and without Au banks along the edges.  The lower graph plots the same data on a log scale showing that oscillations for the device with Au banks are present but greatly suppressed relative to the no banks device.  The addition of banks suppresses the height of the higher-order oscillations consistent with the lateral inverse proximity effect model having an order parameter and maximum critical current density suppressed at the edges and a maximum in between.}
\label{Ic_B}
\end{figure}

Evidence for order parameter suppression near added Au structures by the lateral inverse proximity effect is shown in Fig. \ref{Ic_B}.  In Fig. \ref{Ic_B} we compare measurements of normalized $I_c(B)$ for square $L$ = 16 $\mu$m devices with and without Au banks along the edges.  Adding Au banks to the edges strongly suppresses the $I_c(B)$ side lobe maxima and broadens the central peak. 

\begin{figure}[h]
\includegraphics[width=8cm]{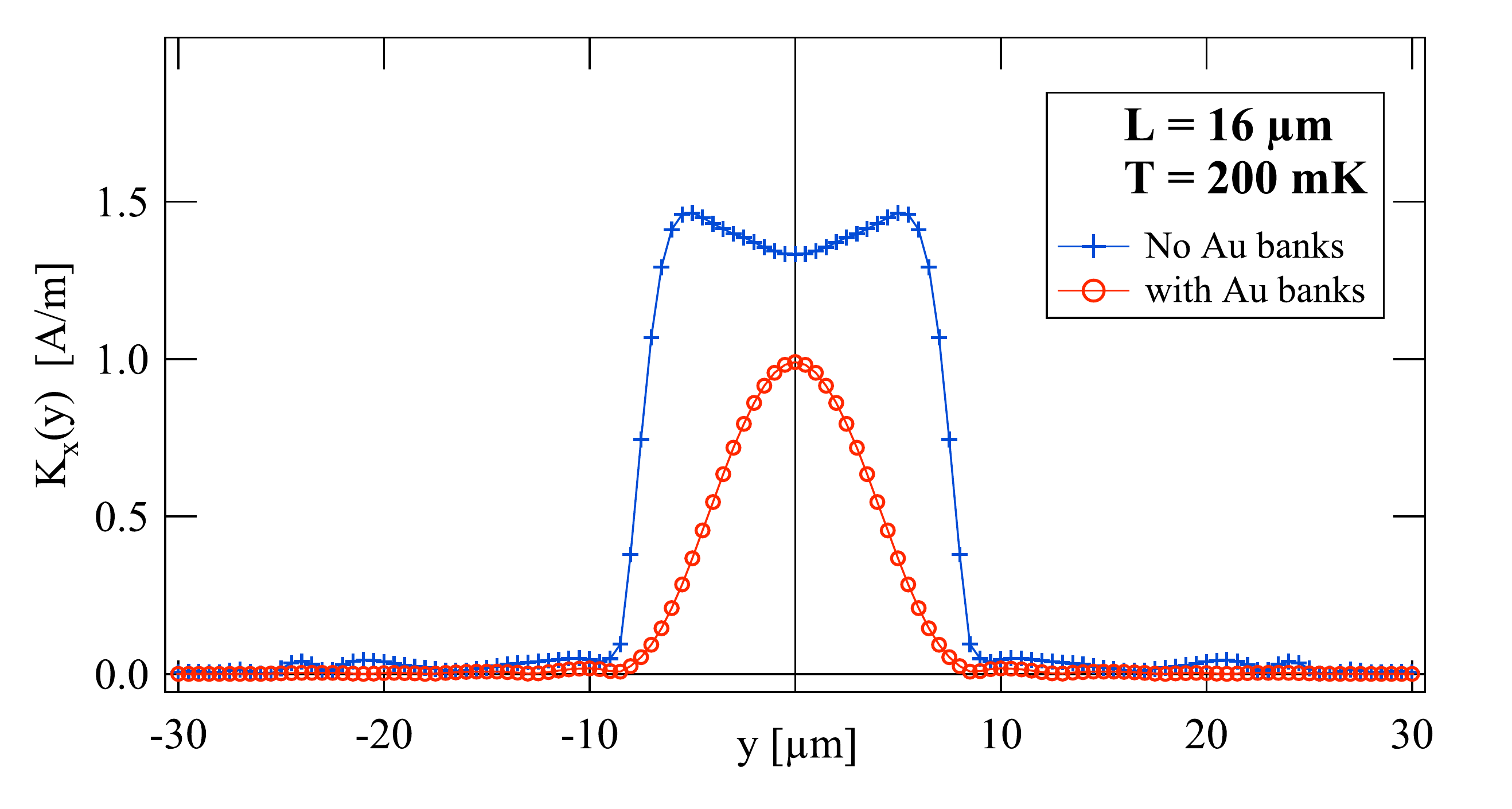}
\caption{%
Calculated $K_x(y)$ versus $y$ for the $L$ = 16 $\mu$m with and without Au banks from the $I_c(B)$ measurements of Fig. \ref{Ic_B} as explained in the text.  The integrated area is larger for the no Au banks as compared to the device with Au banks because the $I_c$ is larger.}
\label{Kx_y}
\end{figure} 

The $I_c(B)$ for a narrow uniformly coupled Josephson junction, for which the field in the weakly linked region is the uniform applied field $\vec B = B \hat z$, follows the well known Fraunhofer pattern given by
\begin{equation}
I_c = I_{c0} \left | \frac{ \sin (\pi \nu )}{\pi \nu} \right |
\end{equation}
where $\nu$ is the magnetic flux in the junction in units of the magnetic flux quantum $\Phi_0$; $\nu \equiv \Phi / \Phi_0$.  When the Josephson coupling is not uniform but instead weakly coupled at the edges and strongly coupled in the middle the resulting $I_c(B)$ pattern changes from the Fraunhofer pattern to having suppressed side lobe maxima and a broader central maxima.  The same changes are expected when applying LaiPE to a TES with Au banks along the edges, as seen in Fig. \ref{Ic_B}.\cite{Sadleir_phd_10}

By symmetry we expect the samples with and without Au banks along the edges to have an even sheet current density $K_x(y)$.  If we then approximate the weak-link as having junction-like local electrodynamics, apply a phase retrieval algorithm, and take the Fourier transform of the $I_c(B)$ data we arrive at the $K_x(y)$ distributions in Fig. \ref{Kx_y}.\cite{Sadleir_phd_10,Carmody99,Carmody00}

In Fig. \ref{Kx_y} we see for the plain square $L$ = 16 $\mu$m Mo/Au bilayer with no banks the $K_x(y)$ approximates a rectangular pulse but with a small dip in the middle and with large but finite sloped edges.  Contrast this with the same size device with Au banks added to the edges and we see a dramatically different $K_x(y)$ that is suppressed at the edges and slowly grows to a maximum in the middle.  This unusual current distribution for a superconducting strip is consistent with the spatially varying order parameter predicted by the LaiPE model that is suppressed at the edges nearest the Au banks and increasing to a maximum $s/2$ away from each edge.  

Critical current versus magnetic field side lobe suppression has been desired for Josephson junction logic and memory applications\cite{Go88,Broom80}, mixers\cite{Reihards79}, optics\cite{Born59}, antennae design\cite{Lo88}, and superconducting tunnel junction particle detectors\cite{Gijsbertsen94}.  Quartic-shaped Josephson junctions have been produced showing suppression of the first side lobe maxima to as low as only a few percent of the zero-field central peak value.\cite{Peterson91,Gijsbertsen94,Broom79}  Instead of tailoring the junction geometry of an SIS structure\cite{Peterson91,Gijsbertsen94} we demonstrate side lobe suppression in SN$^\prime$S weak-links by adding normal-metal layers along the edges.  
In our Josephson coupled structures (a very different geometry than SIS sandwiches) we find suppression of the side lopes to 2\% of the zero-field  central maxima by adding Au banks to the edges.

\section{Nonequilibrium Superconductivity in Devices with Added N-structures}
\label{nonequlibrium_sc_section}

\begin{figure}[h]
\includegraphics[width=8cm]{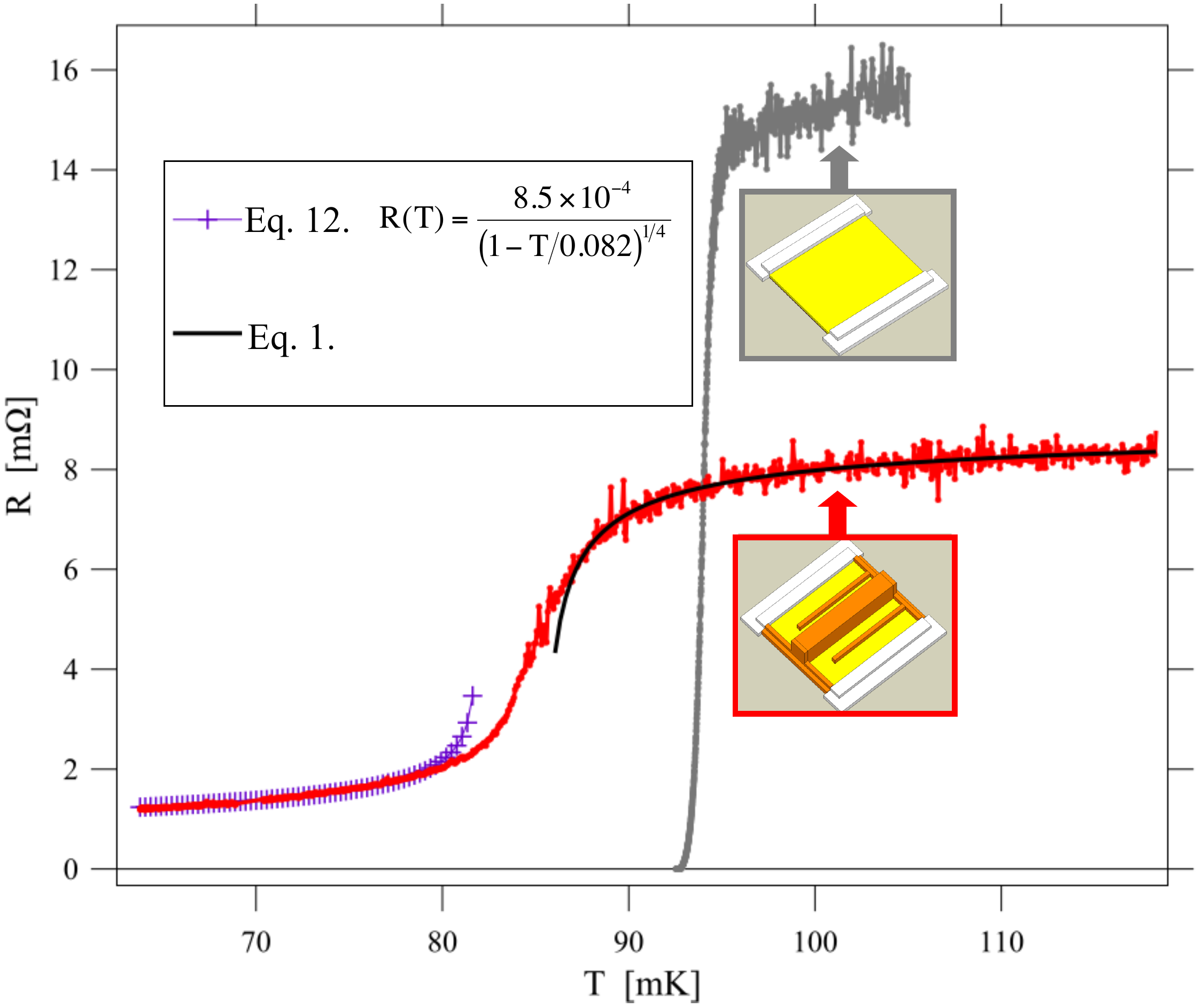}
\caption{%
RT data for $L=110$ $\mu$m devices from the same bilayer with no Au structures and with a Au absorber stem spanning the entire TES width that is 4 $\mu$m thick and 22.5 $\mu$m in the direction of current flow.  Adding Au structures again causes a reduction in $T_c$ and in this case a broad resistive transition and a low-temperature resistive tail, $\sim100$ times larger than the resistance across the Au stem in the normal state.  The high-temperature region is fit to the normal-state slope model for longitudinal proximity effect (Eq. \ref{N_slope}) and the lower-temperature resistive tale is fit to a model using the temperature dependence of the quasiparticle diffusion length (Eq. \ref{RT_lambdaQstar}).  The fit values are in mks units.}
\label{rt_hstem}
\end{figure}

We have also studied TES structures where the added Au layer spans the full width of the TES (Fig. \ref{rt_hstem}).  At low temperatures the current cannot meander around the non-superconducting Mo/Au/Au structures and the current is forced to convert between supercurrent and quasiparticle current.  This conversion processes takes place over a characteristic length scale in the superconductor, $\Lambda_{Q^\ast}$, the quasiparticle diffusion length, given by
\begin{equation}
\Lambda_{Q^\ast}= \sqrt{D\, \tau_Q(T)},
\label{lambdaqstar}
\end{equation}
where $D$ is the electronic diffusion constant and $\tau_Q$ is the charge-imbalance relaxation time.\cite{Tinkham06}   The charge imbalance can relax by various mechanisms and the charge-imbalance relaxation time $\tau_Q$ can be expressed in terms of the BCS superconducting gap $\Delta$ and the electron inelastic scattering time $\tau_E$ as
\begin{equation}
\tau_Q=\frac{4k_BT_c}{\pi \, \Delta} \; \sqrt{ \frac  {\tau_E}  {2 \, \Gamma} },
\label{chargeimbalancetime}
\end{equation}
where
\begin{equation}
\Gamma=
\frac{1}{2\,\tau_E}+
\frac{1}{\tau_s}+
\frac{D\left(\,2\,m\,v_s\right)^2}{2\,\hbar^2}+
\frac{D}{2\, \Delta} \left(   -\frac{\partial^2\Delta}{\partial x^2}  \right),
\label{gamma}
\end{equation}
with the four terms in the $\Gamma$ expression corresponding to inelastic electron-phonon scattering, magnetic impurity spin-flip scattering, elastic scattering from a superfluid current, and gap anisotropy at the S/N interface.  The last three terms may be neglected if we assume no magnetic impurities, $I\ll I_c$, and a slow varying gap--- leaving inelastic electron-phonon scattering as the dominant conversion mechanism.\cite{Schmid75,Yu76,Kadin80}  With the BCS relation for $\Delta_0$=1.76 $k_BT_c$, and $\Delta(T)$ we have, 

\begin{equation}
\Lambda_{Q^\ast}(T)=\Lambda_{Q^\ast}(0)\left(1-\frac{T}{T_c} \right)^{-1/4}
\label{lamdaqstarfit}
\end{equation}
with
\begin{equation}
\Lambda_{Q^\ast}(0)=\left(\tau_Q(0)\, D\right)^{1/2}  \approx  \left(0.723\, \tau_E\, D \right)^{1/2}.
\end{equation}
The expression for the low-temperature dependence of the resistance becomes,
\begin{equation}
R(T)=2\frac{dR}{dx}\Lambda_{Q^\ast}(T)
\label{RT_lambdaQstar}
\end{equation}
Fitting the RT dependence of the resistance tail in Fig. \ \ref{rt_hstem} with the derived temperature dependence of the quasiparticle diffusion length gives a $\tau_Q(0)=1.8\times10^{-10}$ s.  This time is similar to the reported $\tau_Q(0)$ values reported for Sn of $0.9\times10^{-10}$ s and $1.0\times10^{-10}$ s.\cite{Dolan77,Clarke74}

Our understanding for quasiparticle relaxation mechanisms in normal metals at subkelvin temperatures is incomplete and even more so in the case of superconductors.\cite{Timofeev09,Martinis09}  Progress has been made recently in effort to better understand decoherence in Q-bits and performance limitations for other low temperature sensors.\cite{Timofeev09,Martinis09}

The inelastic scattering time $\tau_E$ is not well understood in the superconducting transition nor has it been measured for our Mo/Au films.  Nevertheless, if we compare measurements on Au films and assuming the temperature dependence goes like $T^{-3}$ for a clean normal metal film we may expect a $\tau_E$ value of order $10^{-4}$ s, significantly larger than the $1.3 \times 10^{-10}$ s.  Further investigation is necessary, but this may suggest that other terms in equation \ref{gamma} are not negligible and are playing a significant role in the charge conversion process.  

\section{Bilayer $T_{ci}$ Compensated TES\MakeLowercase{s}}
\label{Tc_compensation_section}

The intrinsic $T_c$ of the Mo/Au bilayer is dependent upon the thickness of each layer and is also sensitive to the tranmissivity of the Mo/Au interface.\cite{Martinis00}  When fabricating Mo/Au bilayers and following identical fabrication procedures (including the same Mo/Au thicknesses measured by atomic force microscopy) it is common for the measured bilayer $T_c$ to change from one bilayer fabrication to the next when identical procedures are followed (\emph{e.g.}, $T_c$ excursions of 25\% are not uncommon).  Each bilayer shows an approximately uniform $T_c$ over its surface and the bilayer $T_c$ tends to rise abruptly upon cleaning the vacuum chamber then drifts downward until the next chamber servicing.  These observations are consistent with the Mo/Au interface transmissivity changing and causing the bilayer $T_c$ variation from one fabrication to the next.  This is why devices with and without added Au structures from the same Mo/Au bilayer are used to determine the $T_c$ shift from the lateral inverse proximity effect for any one data point in Fig. \ref{T_c_s}.  This is also why Figs. \ref{rt_uniformity}, \ref{RT_fingers}, \ref{Ic_T}, \ref{Ic_B}, and \ref{rt_hstem} each compare devices from the same bilayer to remove variation in bilayer $T_c$ as a cause for the differences.

When designing a TES sensor a targeted  bilayer $T_c$ value is chosen.   When a Mo/Au bilayer misses the target $T_c$ value this can make devices unusable or degrade performance.  Incorporating into mask design TES arrays with different $s$ and $L$ spacing, LaiPE and LoPE respectively, can be used to tune the $T_c$ of the TESs therefore compensating for bilayer $T_{ci}$ control fluctuations and ensuring that some arrays from a fabrication run hit the targeted $T_c$ value.\cite{Sadleir08}    Our findings also suggest making TES devices without interface sensitive S/N bilayers by making the TES longitudinally S/N/S where the N material could be a longitudinally proximitized normal-metal (\emph{i.e.}, $T_{ci}=0$), providing much smaller TES thermometer capability.\cite{Sadleir08,Sadleir_prl_10} 

\section{Conclusion}
\label{conclusion}
We have shown that both weak-link and nonequilibrium superconductivity play important roles in TES devices with added normal-metal structures.  We have identified that TES sensors with and without added Au structures exhibit superconducting weak-link behavior over long length scales from measuring the temperature dependence of the critical current and observing Josephson-like oscillations in $I_c$ with applied magnetic field.  As a consequence the transition temperature of a TES is ill defined because it is strongly current dependent and increasingly so as the TES size is reduced.  We find that the strength of the order parameter changes in the plane of the TES film over many tens of $\mu$m.  This is interpreted as a longitudinal proximity effect from the leads and lateral inverse proximity effect from the added N structures. These effects become more pronounced as the superconducting lead separation $L$ and the normal-metal structure separation distance $s$ are reduced. Theoretical attempts to explain the TES transition using fluctuation superconductivity models assume a uniform superconductor and fail to account for the in-plane variations of the average order parameter strength.  By using the measured $I_c(T)$ for a weak-link we can account for the width of the resistive transition in our TESs and presumably in other TESs.

In addition to better understanding the physics of large TESs, many of our findings are vital to the development of TESs of smaller size.  Motivating factors for smaller TES applications include: increased sensitivity to lower energy photons (because of smaller heat capacity), reduced noise in microwave bolometer applications, and developing higher density TES arrays for applications across the EM spectrum.

\begin{acknowledgments}
J.E. Sadleir thanks K.D. Irwin for discussions of nonequilibrium superconductivity, T.M. Klapwijk for discussions of his 2004 publication, and J.R. Clem for many helpful communications.  We also thank J.\ Beyer (PTB Berlin) and K.D. Irwin (NIST Boulder) for providing the SQUIDs used in this work and R. Kelley, F. S. Porter and M. E. Eckart for their support.
\end{acknowledgments}


\begin{thebibliography}{99}
\bibitem{Sadleir_prl_10} J. E. Sadleir, S. J. Smith, S. R. Bandler, J. A. Chervenak, and J. R. Clem, Phys. Rev. Lett., {\bf104}, 047003 (2010). 
\bibitem{IrwinHilton05} K. D. Irwin and G. C. Hilton, in {\it Topics in Applied Physics: Cryogenic Particle Detection}, edited by C. Enss, (Springer, Berlin), (2005).
\bibitem{Bandler08} S. R. Bandler, R. P. Brekosky, A.-D. Brown, J. A. Chervenak, E. Figueroa-Feliciano, F. M. Finkbeiner, N. Iyomoto, R. L. Kelley, C. A. Kilbourne, F. S. Porter, J. Sadleir and S. J. Smith, J. Low Temp. Phys. {\bf 151}, 400 (2008).

\bibitem{Bacrania09} M. K. Bacrania, A. S. Hoover, P. J. Karpius, M. W. Rabin, C. R. Rudy, D. T. Vo, J. A. Beall, D. A. Bennett, W. B. Doriese, G. C. Hilton, R. D. Horansky, K. D. Irwin, N. Jethava, E. Sassi, J. N. Ullom, and L. R. Vale, IEEE Trans on Nuc. Sci, {\bf56}, 2299 (2009).

\bibitem{Ullom04} J. N. Ullom, J. A. Beall, W. B. Doriese, W. D. Duncan, L. Ferreira, G. C. Hilton,
K. D. Irwin, C. D. Reintsema, and L. R. Vale, App. Phys. Lett. {\bf84}, 4206 (2004).

\bibitem{Cuevas07} J. C. Cuevas and F. S. Bergeret, Phys. Rev. Lett. {\bf99}, 217002 (2007).
\bibitem{Heersche07} H. B. Heersche, P. Jarillo-Herrero, J. B. Oostinga, L. M. K Vandersypen, A. F. Morpurgo, Nature {\bf143}, 72 (2007).
\bibitem{leSueur08} H. le Sueur, P. Joyez, H. Pothier, C, Urbina, D. Esteve, Phys. Rev. Lett. 100, 197002 (2008).
\bibitem{Saira07} O.P. Saira, M. Meschke, F. Giazotto, A. M. Savin, M. Mšttšnen, and J. P. Pekola, Phys. Rev. Lett. {\bf99}, 027203 (2007).

\bibitem{Giazotto10} F. Giazotto, J.T. Peltonen, M. Meschke, and J.P. Pekola, Nature Physics {\bf6}, 254 (2010). 
\bibitem{Jarillo-Herrero06}P. Jarillo-Herrero, J.A. van Dam, and L.P. Kouwenhoven, Nature {\bf439}, 953 (2006).

\bibitem{Fraser04} G. W. Fraser, Nucl. Instrum. Meth. A {\bf 523}, 234 (2004).
\bibitem{Seidel04} G. M. Seidel and I. S. Beloborodov, Nucl. Instrum. Meth. A {\bf 520},  325 (2004).
\bibitem{LindemanPerc06} M.A. Lindeman, M.B. Anderson, S.R. Bandler, N. Bilgri, J. Chervenak, S.G. Crowder, S. Fallows, E. Figueroa-Feliciano, F. Finkbeiner, N. Iyomoto, R. Kelley, C.A. Kilbourne, T. Lai, J. Man, D. McCammon, K.L. Nelms, F.S. Porter, L.E. Rocks, T. Saab  J. Sadleir, and G. Vidugiris, Nucl. Instrum. Meth. A {\bf559}, 715 (2006).

\bibitem{Brandt09} D Brandt, AIP Conf. Proc. {\bf1185}, 52 (2009).

\bibitem{Luukanen03} A. Luukanen, K. M. Kinnunen, A. K. NuottajŠrvi, H. F. C. Hoevers, W. M. Bergmann Tiest, and J. P. Pekola, Phys. Rev. Lett. {\bf90}, 238306 (2003).


\bibitem{Buzdin05} A.I. Buzdin, Rev. Mod. Phys. 77, 935Ð976 (2005).

\bibitem{Martinis00} J. M. Martinis, G. C. Hilton, K. D. Irwin, and D. A. Wollman, Nucl. Instrum. Meth. A {\bf 444}, 23 (2000).  This reference is used to estimate the trilayer $T_c$ assuming an interface transmission coefficient for the Au/Au interface that is the same as the Au/Mo interface.
\bibitem{Cretinon04} L. Cretinon, A. Gupta, B. Pannetier, and H. Courtois, Physica C {\bf  404} 103 (2004). 
\bibitem{Moussy01}N. Moussy, H. Courtois, and B. Pannetier, Rev. Sci Instru {\bf 72} 128 (2001).  
\bibitem{Tessmer96} S. H. Tessmer, M. B. Tarlie, D. J. Van Harlingen, D. L. Maslov, and P. M. Goldbart, Phy. Rev. Lett. {\bf 77} 924 (1996).
\bibitem{Sadleir_phd_10} J. E. Sadleir, PhD. Dissertation, University of Illinois Physics Dept. (2010). 
\bibitem{stress_comment} The measured $T_c$ shifts and scaling found for devices with different Au structure designs can not be explained in terms of the added Au structures altering the film stress in the bilayer and shifting $T_c$.  The measured size of the shift is larger than stress estimates; in addition, TESs measured on solid substrates and on thin perforated Si$_3$N$_4$ membranes gave the same transition.

\bibitem{Chervenak04} J. A. Chervenak, F. M.  Finkbeiner, T. R. Stevenson, D. J. Talley, R. P. Brekosky, S. R. Bandler, E. Figueroa-Feliciano, M. A. Lindeman, R. L. Kelley, T. Saab, and C. K. Stahle, Nucl. Instrum. Meth. A, {\bf 520}, 460 (2004).
\bibitem{Smith08} S. J. Smith, S. R. Bandler, R. P. Brekosky, A. -D. Brown, J. A. Chervenak, F. M. Finkbeiner, N. Iyomoto, R. L. Kelley, C. A. Kilbourne, F. S. Porter, and J. E. Sadleir,  J. Low Temp. Phys., {\bf151}, 195 (2008).
\bibitem{Rn3finger}  In Fig. \ref{RT_fingers}  6 and 9 finger devices have $L = 130$ $\mu$m whereas the $R_N$ value is lower for the 3 finger device because for this sample $L = 110$ $\mu$m.  This agrees with the resistor network model using the measured resistivities and geometry for each element.
\bibitem{Glover67} R.E. Glover, Phys. Rev. Lett., {\bf25}, 542 (1967).
\bibitem{Ginzburg60} V.L. Ginzburg, Sov. Solid State Phys. {\bf2}, 61 (1960).
\bibitem{Aslamazov68} L.G. Aslamazov, and A.I. Larkin, Sov. Solid State Phys. {\bf10}, 875 (1968).
\bibitem{Maki68} K. Maki, Prog. in Theor.Phys. {\bf39}, 897 (1968).

\bibitem{Tinkham06} M. Tinkham, {\it Introduction to Superconductivity}, 2nd Ed., McGraw-Hill, NY, 1996.
\bibitem{DeGennes99} P. G. De Gennes,  {\it Superconductivity of Metals and Alloys}, 2nd Ed., Westview Press, 1999.
\bibitem{Skopol75} W. J. Skopol, M. Tinkham Rep. Prog. Phys. {\bf38}, 1049 (1975).

\bibitem{Sadleir08} J. E. Sadleir, NASA New Tech. Rep.: {\it Novel Superconducting Transition Edge Sensor Design}, 5026629 (2008).

\bibitem{Liniger93} W. Liniger, J. Low Temp. Phys. {\bf 93}, 1 (1993).

\bibitem{Boogaard04}G. R. Boogaard, A. H. Verbruggen, W. Belzig, and T. M. Klapwijk,  Phys. Rev. B  {\bf69}, 2205031 (2004).

\bibitem{Carmody99} M. Carmody, E. Landree, L. D. Marks, K. L. Merkle, Physica C {\bf315} 145 (1999).
\bibitem{Carmody00} M. Carmody, B. H. Moeckly, K. L. Merkle, and L. D. Marks, J. Appl. Phys. {\bf87} 2454 (2000).

\bibitem{Go88}D. Go, C.A. Hamilton, F.L. Lloyd, M.S. Dilorio, and R.S. Withers, IEEE Trans Electron Devices  {\bf35}, 498 (1988).
\bibitem{Broom80}R.F. Broom, W. Kotyczka, and A. Moser, IBM J. Res. Dev. {\bf24}, 178 (1980).
\bibitem{Reihards79} P.L. Riehards, T.M. Shen, R.E. Harris, and F.L. Lloyd, Appl. Phys. Lett. {\bf34}, 345 (1979).
\bibitem{Born59}Born, M. and Wolf, E. Principles of Optics Pergamon Press, New York  416 (1959).
\bibitem{Lo88}Y.T. Lo, and S.W. Lee,  (Eds) Antenna Handbook Van Nostrand Reinhold, New York (1988).
\bibitem{Broom79}R.F. Broom, P. Gueret, W. Kotyczka, T.O. Mohr, A. Moser, A. Oosenbrng, and  P. Wolf, IEEE J Solid-State Circuits, SC-14, 690 (1979).
\bibitem{Peterson91}R.L. Peterson, Cryogenics {\bf31} 132 (1991).
\bibitem{Gijsbertsen94} J. G. Gijsbertsen, E. P. Houwman, B. B. G. Klopman, J. Flokstra, H. Rogalla, D. Quenter,  and S. Lemke, Physica C {\bf249} 12 (1995).


\bibitem{Schmid75} A. Schmid, G. Schon, J. Low Temp. Phys. {\bf20} 207 (1975).
\bibitem{Yu76} M. L. Yu and J. E. Mercereau, Phys. Rev. B {\bf12} 4909 (1976).

\bibitem{Kadin80} A. M. Kadin, L.N. Smith, and W. J. Skocpol J. Low Temp Phys {\bf 38} 497 (1980).

\bibitem{Dolan77} G. J. Dolan and L. D. Jackel Phys. Rev. Lett. {\bf39}, 1631 (1977).
\bibitem{Clarke74} J. Clarke and J. Patterson, J. Low Temp. Phys. {\bf15}, 491 (1974).
\bibitem{Timofeev09} A. V. Timofeev, C. Pascual Garcia, N. B. Kopnin, A. M. Savin, M. Meschke, F. Giazotto, and J. P. Pekola, Phys. Rev. Lett. {\bf102} 017003 (2009).
\bibitem{Martinis09} John M. Martinis, M. Ansmann, and J. Aumentado, Phys. Rev. Lett. {\bf103} 097002 (2009).

\end{thebibliography}
\end{document}